\documentclass{aipproc} 

\usepackage[english]{babel}
\usepackage[latin1]{inputenc}

\layoutstyle{6s}

\newcommand{\qu}{$u$}
\newcommand{\qd}{$d$}
\newcommand{\qs}{$s$}
\newcommand{\qc}{$c$}
\newcommand{\qb}{$b$}
\newcommand{\jp}[1]{\mbox{$\frac{#1}{2}\mbox{}^+$}}
\newcommand{\spl}[1]{\mbox{$#1^+$}}

\renewcommand{\indent}{\hspace{\parindent}}

\begin{document}

\author{C. Albertus}{address={Dpto. de Física Moderna, U. Granada, Spain.}}
\author{J. E. Amaro}{address={Dpto. de Física Moderna, U. Granada, Spain.}}
\author{E. Hernández}{address={Grupo de Física Nuclear, Facultad de Ciencias, U. Salamanca, Spain.}}  
\author{J. Nieves}{address={Dpto. de Física Moderna, U. Granada, Spain.}}  

\title{Charm- and Bottom- Baryons: A Variational Approach 
Using Heavy Quark Symmetry} 

\begin{abstract}
\rule{0ex}{3ex}We evaluate masses  of bottom and charmed baryons 
using several non-relativistic quark potentials which parameters have
been adjusted to the meson spectra. Heavy Quark Symmetry leads to
important simplifications of the three body problem, which turns out
to be easily solved by a simple variational ansatz. The wave functions
obtained can be readily used to compute further observables as mass
densities or form factors. The quark-quark potentials explored so far,
show an overall good agreement with the experimental masses.
\end{abstract}

\maketitle

\section{Introduction}
\indent The non-relativistic constituent quark model (NRCQM), using
QCD-inspired potentials, has proved to be an excellent tool to predict
properties of hadrons.

In the case of baryons including one heavy flavour (\qc, \qb) and two
light ones (\qu, \qd, \qs), it is possible to take advantage of yet
another property of QCD: Heavy Quark Symmetry (HQS)
Ref.~\cite{iw89,neu94}. In the limit in which the mass of the heavy
quark is infinity, the quantum numbers of the light degrees of freedom
are well defined always. Furthermore, in this limit, the masses of the
baryons depend only on the quark content and on the light-light
quantum numbers of the baryon. All of this is a clear simplification
for solving the three body problem.
%
%
%
%
Thus, for bottom- and charm-baryons we can consider the quantum
numbers of the two light quark system to be fixed, and neglect
corrections terms in the wave function that scale as
${\cal O}\left(\Lambda_{QCD}/m_{c,b}\right)$.

The aim of this work is to determine masses and other properties like
mass densities and electromagnetic form factors for baryons containing
a heavy quark and two light ones. This study includes all baryons
compiled in Table \ref{tab:bariones} and some more details will be
given elsewhere \cite{aahn03}.
\smallskip
\begin{table}[h]
\begin{tabular}{cccccc|cccccc}
\hline
Baryon           & (S) & $J^P$  & (I)             & $s_l^{\pi_l}$& 
Quark content    & Baryon           & (S) & $J^P$  & (I)    & $s_l^{\pi_l}$& 
Quark content     \\

\hline
$\Lambda_{c,b}$  & (0) & \jp{1} & (0)             & \spl{0}      &
 (\qu,\qd)\qc,\qb 
& $\Xi_{c,b}'$     &(-1) & \jp{1} & ($\frac{1}{2}$) & \spl{1}      &
 (\qu,\qs)\qc,\qb  \\

$\Sigma_{c,b}$   & (0) & \jp{1} & (1)             & \spl{1}      &
 (\qu,\qu)\qc,\qb  
& $\Xi_{c,b}^*$    &(-1) & \jp{3} & ($\frac{1}{2}$) & \spl{1}      &
 (\qu,\qs)\qc,\qb  \\

$\Sigma_{c,b}^*$ & (0) & \jp{3} & (1)             & \spl{1}      &
 (\qu,\qu)\qc,\qb  
& $\Omega_{c,b}$   &(-2) & \jp{1} & (0)             & \spl{1}      &
 (\qs,\qs)\qc,\qb  \\

$\Xi_{c,b}$      &(-1) & \jp{1} & ($\frac{1}{2}$) & \spl{0}      &
 (\qu,\qs)\qc,\qb  
&$\Omega_{c,b}^*$ &(-2) & \jp{3} & (0)             & \spl{1}      &
 (\qs,\qs)\qc,\qb  \\
\hline
\end{tabular}
\label{tab:bariones}
\caption{Baryons considered in this work. The information enclosed in
the different columns is strangeness, spin-parity, isospin,
spin-parity of the light degrees of freedom and quark content. The
spin-parity of the light quarks, fifth and eleventh columns, in some
cases are determined thanks to HQS.}
\end{table}

\section{ The Three-Body Problem}

\indent The intrinsic hamiltonian that describes the dynamics
of the baryon is given by\footnote{In this hamiltonian the motion of
the center of mass (CM) of the baryon has been taken out.}
\begin{equation}
H=-\frac{\vec\nabla_1^2}{2\mu_{1}}-\frac{\vec\nabla_2^2}{2\mu_{2}}+
\frac{\vec\nabla_1\cdot\vec\nabla_2}{m_{h}}+
V_{l_1h}(\vec{r}_1,spin)+V_{l_2h}(\vec{r}_2,spin)+
V_{l_1l_2}(\vec{r}_1,\vec{r}_2,spin),
\label{eq:ham}
\end{equation}
where $\vec r_i$ is the position of the $i$-th light quark with
respect to the heavy one, $m_h$ stands for the mass of the heavy
quark, while $\mu_i$ accounts the reduced mass of the heavy and the
$i$-th light quark system, $V_{l_ih}$ and $V_{l_1l_2}$ are the
light--heavy and light--light interaction potentials, and the words
$spin$ stands for possible spin dependence of the potentials. Note the
presence of the Hughes-Eckart term
$\vec\nabla_1\cdot\vec\nabla_2/m_{h}$ that results from the separation
of the CM movement.

The potentials used in this work are the one proposed by R.K. Bhaduri
et al. in Ref.~\cite{bha81}, and the set of potentials proposed by
B. Silvestre-Brac and C. Semay that can be found in Ref.~\cite{sil96}.
The parameters of those potentials have been adjusted in the meson
sector.  For their use in the $qq$ sector they have to be adequately
transformed. We use the prescription $V_{ij}^{qq}=\frac{1}{2}V_{ij}^{q
\bar q}$ that assumes a $\vec\lambda_i \cdot \vec\lambda_j$ color
dependence in all terms of the potential \cite{sil96}.


To solve the three-body problem one can use Faddeev equations
\cite{sil96}. This is a non-trivial task from the computational point
of view, and leads to wave functions that are difficult to use in
other contexts. Here we propose an extremely simple variational
scheme. As it is usual, we assume an antisymmetric wave-function for
the color degrees of freedom and the spin-flavour wave function is
determined by the quantum numbers specified in
Table~\ref{tab:bariones}. Finally for the spatial wave function, we
propose the ansatz
%
\begin{equation}
\psi(r_1,r_2,r_{12})=NF(r_{12})\phi_1(r_1)\phi_2(r_2)
\end{equation}
where $N$ is a normalization factor, $\phi_i(r_i)$ is the ground state
wave function for the $V^{qq}_{l_i\,h}$ potential, and $F(r_{12})$ is
a Jastrow correlation function in the relative distance of the two
light quarks $r_{12}$\footnote{We have assumed that the relative
orbital angular momentum between the light quarks is zero. Thus the
spatial wave function can only depend on $r_1$, $r_2$ and $r_{12}$}. For
$F$ we take
\begin{equation}
F(r_{12})=\left(1-e^{-c_1 r_{12}}\right)
\sum_{j=2}^{4}a_{j}e^{-b^2_{j}(r_{12}-d_{j})^2}
\end{equation}
\noindent where the term $e^{-c_1 r_{12}}$ would be excluded 
in those cases where the potential $V_{ll}^{qq}$ does not show a
repulsive hard core at the origin.  Taking into account that the color
wave function is antisymmetric we use symmetrized wave functions in
the spin-isospin and orbital degrees of freedom of the two light
quarks.

This variational scheme shows clear resemblances to that succesfully
used in the study of double $\Lambda$ hypernuclei \cite{aan02}.

\section{Preliminary Results}
\indent Our results for the masses obtained with the AL1 potential of
Ref.~\cite{sil96} are given in Tables \ref{tab:b-bariones} and
\ref{tab:c-bariones}.  We find good agreement with experimental data
\cite{hag02}, when available, and with previous results from lattice
\cite{bow96} and Faddeev calculations \cite{sil96}. 


In Table \ref{tab:radios} we give the mass radii obtained also for the
AL1 potential. Our results agree with the ones obtained in
Ref.~\cite{sil96} using a Faddeev approach. Conclusions are similar
when the potential of \cite{bha81} or potentials AL2, AP1 or AP2 of
Ref.~\cite{sil96} are used.

\smallskip
%
\begin{table}[h]
\begin{tabular}{ccccccc}\hline
B &$s^\pi$& content & $M_{exp.}$ [MeV] & $M_{Latt.}$ [MeV]& $M_{Var}$
[MeV] & $M_{Fad.}$ [MeV] \\\hline $\Lambda_b$&$0^+$&$udb$& $ 5624 \pm
9$ & $5640 \pm 60 $ & $5640$ & $5638$ \\ $\Sigma_b$ &$1^+$&$llb$& &
$5770 \pm 70 $ & $5846$ & $5845$ \\ $\Sigma^*_b$ &$1^+$&$llb$& & $5780
\pm 70$ & $5877$ & \\ $\Xi_b$ &$0^+$&$lsb$& & $5760 \pm 60 $ & $5805$
& $5806$\\ $\Xi'_b$ &$1^+$&$lsb$& & $5900 \pm 70$ & $5941$ & \\
$\Xi^*_b$&$1^+$&$lsb$& & $5900 \pm 80 $ & $5972$ &\\ $\Omega_b$
&$1^+$&$ssb$& & $5990 \pm 70 $ & $6034$ & $6034$ \\ $\Omega^*_b$
&$1^+$&$ssb$& & $6000 \pm 70$& $6065$ & \\ \hline
\end{tabular}
\caption{ Masses for the bottom- baryons considered. The
spin-parity of the light degrees of freedom is shown in the second
column. Results with our variational approach and with a Faddeev
calculation from Ref.~\cite{sil96} are included. Lattice
QCD~\cite{bow96} and experimental values~\cite{hag02}, when available,
are also given.}
\label{tab:b-bariones}
\end{table}
\begin{table}[h]
\begin{tabular}{ccccccc}\hline
B &$s^\pi$& content & $M_{exp.}$ [MeV]&
$M_{Latt.}$ [MeV]& $M_{Var}$ [MeV]& $M_{Fad.}$ [MeV]\\\hline 
$\Lambda_c$&$0^+$&$udc$& $2285 \pm 1$ & $2270 \pm 50 $ & $2291$ & $2285$\\ 
$\Sigma_c$ & $ 1^+$&$llc$& $2452 \pm 1$ & $2460 \pm 80 $ & $2453$ & $2455$\\
$\Sigma^*_c$ &$1^+$&$llc$& $2518\pm 2$ &$ 2440 \pm 70 $ & $2542$ & \\ 
$\Xi_c$ &$0^+$&$lsc$& $2469 \pm 3$ & $2410 \pm 50 $ & $2476$ & $2467$\\ 
$\Xi'_c$ &$1^+$&$lsc$& $2576\pm 2$ & $2570 \pm 80$ & $2571$ & \\ 
$\Xi^*_c$ &$1^+$&$lsc$& $2646 \pm 2$ & $2550 \pm 80$ & $2657$ & \\
$\Omega_c$ &$1^+$&$ssc$& $2698 \pm 3$ & $2680 \pm 70$ & $2677$ &$2675$ \\ 
$\Omega^*_c$ &$1^+$&$ssc$& & $2660 \pm 80$ & $2761$ & \\\hline 
\end{tabular}
\label{tab:c-bariones}   
\caption{As in Table~\ref{tab:b-bariones} for the charm sector.}
\end{table}

\begin{table}[h!]
\begin{tabular}{ccc}
\hline
B            & $\langle r^2\rangle\ [fm^2] $(Var) & 
$\langle r^2\rangle[fm^2]$ (Fad.) \\
\hline
$\Lambda_b$  &   0.045          &    0.045         \\
$\Sigma_b$   &   0.054          &    0.054         \\
$\Xi_b$      &   0.048          &    0.048         \\
$\Omega_b$   &   0.054          &    0.054         \\
$\Lambda_c$  &   0.095          &    0.104         \\
$\Sigma_c$   &   0.117          &    0.121         \\
$\Xi_c$      &   0.096          &    0.104         \\
$\Omega_c$   &   0.102          &    0.108         \\
\hline
\end{tabular}
\caption{Results for mass radii using this variational 
here and those from the Faddeev calculation of Ref.~\cite{sil96}.}%
\label{tab:radios}
\end{table}

\begin{figure}
\includegraphics{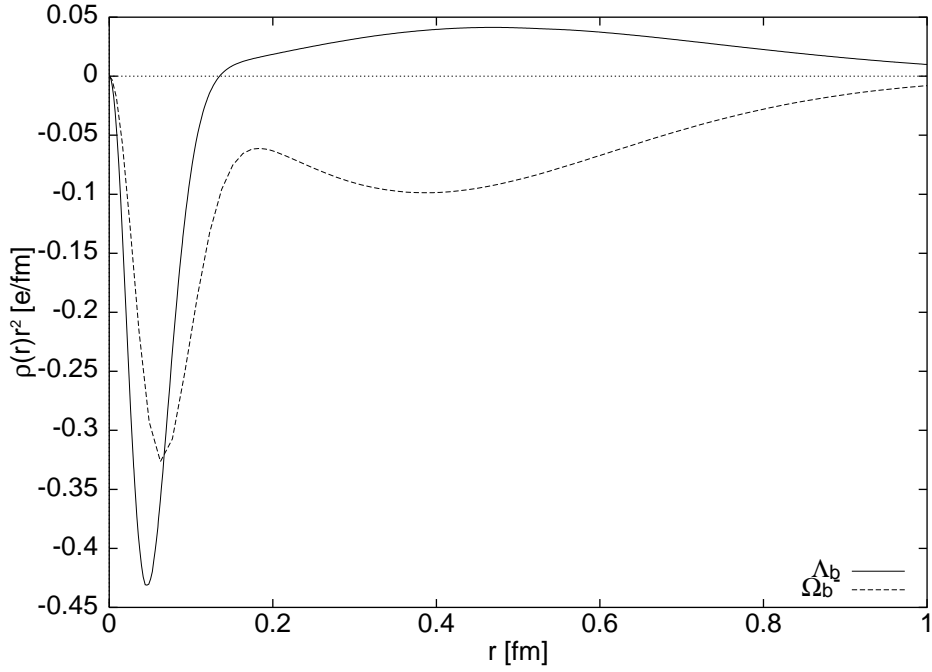}
\label{fig:1}
\caption{Charge density times $r^2$ for 
$\Lambda_b$ (solid) and $\Omega_b^-$ (dashed).}
\end{figure}

\begin{figure}
\includegraphics{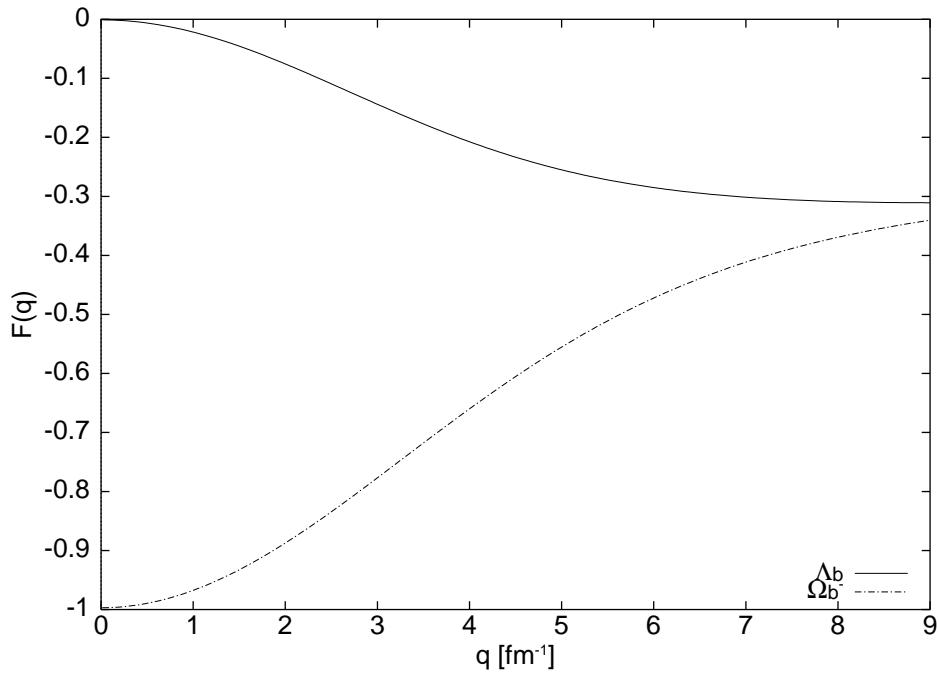}
\label{fig:2}
\caption{Electric form factors for $\Lambda_b$ (solid) and $\Omega_b$ 
(dot-dashed). 
The value at the origin is the charge of the baryon, 0 for $\Lambda_b$
and -1 for $\Omega_b^-$}
\end{figure}

Finally in Figs. \ref{fig:1} and \ref{fig:2} we give our results for the 
charge density and electric form factor of the 
 $\Lambda_b$ and $\Omega_b^-$ baryons. 
\section{Concluding Remarks}
The use of HQS simplifies considerably the solution of the three body
 problem in baryons with a heavy quark. Here we propose a method based
 on a simple variational approach that provides us with simple and
 portable wave functions that can be used in other contexts.  Our
 results agree with previous ones obtained in the lattice or using a
 more complicate Faddeev approach.  Calculations with potentials
 obtained from chiral quark models \cite{bla99} and the study of the
 semileptonic decay of bottom baryons into charmed ones are under
 consideration.

\section{Acknowledgments}
\indent This work is supported by Spanish DGICYT and FEDER funds, 
under contracts no.  BFM2000-1326 and BFM2002-03218 and by Junta de
Andalucía and Junta de Castilla y León under contracts no. FQM225 and
no. SA109/01. C. A. wishes to acknowledge a PhD. grant from Junta de
Andalucía.


\end{document}